# Exoplanets – search methods, discoveries, and prospects for astrobiology

**B. W. JONES**

*Astronomy Group, Physics & Astronomy, The Open University, Milton Keynes, MK7 6AA, UK*

*e-mail: b.w.jones@open.ac.uk   Phone: +44 1908 653229        Fax: +44 1908 654192*

**(SHORT TITLE: Exoplanets and astrobiology)**

**Abstract**

Whereas the Solar System has Mars and Europa as the best candidates for finding fossil/extant life as we know it – based on complex carbon compounds and liquid water – the 263 (non-pulsar) planetary systems around other stars known at 15 September 2008 could between them possess many more planets where life might exist. Moreover, the number of these exoplanetary systems is growing steadily, and with this growth there is an increase in the number of planets that could bear carbon-liquid water life. In this brief review the main methods by which exoplanets are being discovered are outlined, and then the discoveries that have so far been made are presented. This is followed by an account of likely future discoveries. Habitability is then discussed, and an outline presented of how a planet could be studied from afar to determine whether it is habitable, and whether it is indeed inhabited. This review is aimed at the astrobiology community, which spans many disciplines, few of which involve exoplanets. It is therefore at a basic level and concentrates on the major topics.

**Introduction**

The Solar System has Mars and Europa as the best candidates for finding fossil/extant life as we know it – based on complex carbon compounds and liquid water. However, the 263 planetary systems known around other stars at 15 September 2008 greatly increases the number of planets where life might exist (this excludes four planets orbiting pulsars – see below). There is a steady growth in the number of known exoplanetary systems, which is steadily increasing the number of planets that could bear carbon-liquid water life. Clearly, to increase our chance of finding extraterrestrial life we must look beyond the Solar System. In this brief review I will outline the main methods by which exoplanets are being discovered, the discoveries that have been made, and likely future discoveries. I will then outline habitability, how a planet could be studied from afar to determine whether it is habitable, and whether it is in fact inhabited. This review is aimed at the astrobiology community, which spans many disciplines, few of which involve exoplanets. It is therefore at a basic level and concentrates on the major topics. It is divided into four parts:

    how to find exoplanets, and what we can learn about them

    the known exoplanets

    the unknown exoplanets

habitable planets

how to determine from afar whether an exoplanet is habitable/inhabited.

**How to find exoplanets, and what we can learn about them**

The most obvious way to find an exoplanet might seem to be to scrutinise stars with a large telescope, to see if any exoplanets are visible. Alas! this *direct method* is barely feasible at present even under the most favourable conditions. This is because the distance from a star to a planet orbiting it is very much smaller than the distance to the star from us. Therefore the angular separation between star and planet is tiny. Moreover, the planet is the order of a billion times fainter than its star at visible wavelengths, and a million times fainter at infrared wavelengths. Also, all telescopes blur images due to a fundamental optical limit (the diffraction limit). This means that the bright stellar image overlaps the faint planetary image, completely obscuring it. The larger the telescope the smaller the blur, but even large telescopes, with apertures of many metres diameter, can only see really large planets well separated from their star. This has been achieved in only a few cases.

We therefore use indirect methods to detect exoplanets, methods depending on the planet's effect on its star's motion through space or on the light we receive from its star. There are four main methods:

    astrometry

    radial velocity measurements

    transit photometry

    gravitational microlensing.

I present brief outlines of each method: for further details consult Jones (2004 Chapters 9 & 10, 2008 Chapters 8-10), also Perryman (2000), and Pudritz et al. (2007 Chapters 1 & 2).

*Astrometry*

This relies on the motion of the star around the centre of mass of the star-planet(s) system. For a planet of mass $m$ and a star of mass $M$

$$M \times a_S = m \times a_p \tag{1}$$

where $a_S$ and $a_p$ are, respectively, the semimajor axis of the star's and planet's orbits with respect to the centre of mass. By applying Newton's laws of motion and law of gravity it can be shown that, if $M \gg m$

$$\beta \approx [G/(4\pi^2)]^{1/3} [P/M]^{2/3} m/d \qquad (2)$$

where $\beta$ is the angular size of the star's orbit (in radians), $G$ is the universal constant of gravity, $P$ is the orbital period of the star (and of the planet), and $d$ is the distance of the star from us.

Figure 1 shows the motion of the Sun's centre around the centre of mass of the Solar System, viewed face-on from a distance of 30 light years. The dashed circular line is the path the Sun's centre would take if Jupiter were the only planet in the Solar System – you can see that the diameter of this path is only a little greater than the diameter of the Sun. The other path is the Sun's actual motion, due to all of the planets in the Solar System. A planet's influence increases as its mass increases, and as its distance from the Sun increases, and therefore as its orbital period increases (equation 2). Jupiter, with a mass 317.8 times that of the Earth, and an orbit with an orbital period of 11.86 years, dominates. Saturn, 95.2 Earth masses ($m_E$), orbital period 29.42 years, has the second greatest influence. Equation (2) shows clearly the Earth's contribution is tiny.

> Figure 1   The motion of the Sun around the centre of mass of the Solar System, viewed face-on from a distance of 30 light years. The dashed circular line is the path the Sun's centre would take if Jupiter were the only planet in the Solar System. The other path is the Sun's actual motion, due to all of the planets in the Solar System. (The Figures are grouped at the end.)

Note that were the orbit not presented face-on, the path of the Sun on the sky would be different, but the maximum excursions would be the same.

Within 30 light years[1] there are about 500 stars, roughly 70% of which are in multiple systems, mainly binary stars. Our Galaxy is about 100 000 light years across, and contains about 200 thousand million stars, and so 30 light years is well within our cosmic back yard. Even so, the angular motions in Figure 1 are tiny, spanning a little under 0.002 arcsec[2] i.e. 2 milli-arcsec. This is to be compared with the Moon's angular diameter of about 1800 arcsec! Such tiny angles are extremely difficult to measure, and to date no exoplanets have been discovered by astrometry. This is unfortunate, because from astrometry we would get the mass of the planet (if we know the mass of the star, which can usually be obtained from various stellar observations), its orbital period, semimajor axis, and its eccentricity (see Table 1 below).

---

[1] A light year is the distance that light travels through a vacuum in one year. It equals $9.460536 \times 10^{12}$ km, which is 63239.8 AU, the AU being the semimajor axis of the Earth's orbit around the Sun.
[2] An arcsec is 1/3600 of a degree of arc.

What of the future? From the ground, by a technique called interferometry, whereby the output sfrom two or more telescopes are combined, very small angular motions can be detected. For example, at the European Southern Observatory's Very Large Telescope (actually four large telescopes, each 8.2 metres aperture), in a project called PRIMA (PRIMA 2007), angular motions as small as the order of 10 micro-arcsec should be detectable. It is about to start operations. In space, the European Space Agency's satellite Hipparcos, in operation from August 1989 to March 1993, detected position changes down to about one milli-arcsec (Hipparcos 2007), and the forthcoming Gaia satellite (ESA 2008) and NASA's SIMPlanetQuest (PlanetQuest 2008) will get down to about a micro-arcsec.

*Radial velocity measurements*

This technique has yielded the great majority of the 300 or so exoplanet discoveries. It depends on the shift in the wavelength of radiation received from a source when it has a component of motion along the radial direction to the observer. If the motion is towards the observer the received wavelengths are shorter than those emitted by the source, and longer if the motion is away from the observer. This shift is called the Doppler effect. Whereas for the astrometric method the orientation of the star's orbit is immaterial, in the radial velocity method the orientation has to be other than face-on, otherwise there is no variation in the radial velocity.

In the case of a star, as well as a constant drift through space, there will be a cyclic radial motion if the star is orbiting the centre of mass of a planetary system (provided that the orbit of the star is not face-on). It is the spectral absorption lines of a star that have well defined wavelengths at the star, and it is the received wavelengths (particularly at visible and near infrared wavelengths) that enable us to obtain information about the planet's orbit and mass, in accord with

$$v \approx m \, (2\pi G)^{1/3} / (P^{1/3} M^{2/3}) \tag{3}$$

for a circular orbit presented edge-on, where $v$ is the amplitude of the (sinusoidal) radial velocity, and is also the speed of the star around its orbit.

If the orbit is not presented edge-on then in the radial direction we get the component $v \sin(i)$ of the orbital speed where $i$ is the inclination angle shown in Figure 2. The observed amplitude is thus

$$v_{rA} = v \sin(i) \tag{4}$$

It then follows, from equations (3) and (4) that

$$m \sin(i) = v_{rA} \, (P^{1/3} M^{2/3})/(2\pi G)^{1/3} \tag{5}$$

Unless $i = 90°$ (edge-on), the measurement of $v_{rA}$ gives $m \sin(i)$, not $m$, so the actual planetary mass $m$ will then be greater than that indicated by our measurements. This would not be a problem if we knew $i$. Unfortunately this is not often the case. The consequences, and the exceptions, are discussed later.

Figure 2    The inclination angle ($i$ in the text).

So far I have considered circular orbits. For non-circular orbits the speed varies around the orbit, being greatest when the star is closest to the centre of mass, the greater the orbital eccentricity the greater the speed[3]. The radial velocity $v_r$ is then non-sinusoidal around the orbit. It is also the case that, for a given semimajor axis of the planet with respect to the star, $v_{rA}$ is greater the higher the eccentricity. This favours the detection of high eccentricity orbits.

Figure 3 gives some early examples of radial velocity measurements. The upper data are for a low eccentricity orbit. The other two sets of data are for eccentric orbits of different orientation with respect to the radial direction. Note how modest the radial velocities of the stars are. In all cases we can determine the eccentricity of the planet's orbit, and also the value of $m \sin(i)$. The details will not concern us. If more than one planet is present the radial velocity curves are more complex, but the same information can be obtained as for the case of a single planet (e.g. Vogt et al. 2005). Again, the details will not concern us.

Figure 3    Early radial velocity of three stars. Upper curve – 51 Pegasi, near-circular orbit, eccentricity 0.0197. Middle curve – 70 Virginis, orbital eccentricity 0.40. Lower curve – 16 Cygni B, orbital eccentricity 0.689, differently oriented towards us than 70 Viginis.

Detection of changes in $v_r$ as small as about a metre per second is now routinely made, and equipment is becoming available that will get down to 0.1-0.01 m s$^{-1}$. Clearly this will improve our ability to detect exoplanets.

*Transit photometry*

When an exoplanet's orbit is presented to us sufficiently close to edge-on, then the planet will pass between us and its star. This results in an apparent decrease in the brightness of the star, as illustrated in Figure 4. A local example of this phenomenon was the transit of Venus across the face of the Sun on 08 June 2004. The next such transit will be on 06 June 2012, but then not again until 2117.

---

[3] The orbital eccentricity is a measure of the departure of an orbit from circular form. The eccentricity of a circular orbit is zero. Only at values above about 0.2 is the departure from circular form readily apparent. An eccentricity of 0.99 corresponds to a highly elongated orbit not very different from a straight line.

If the star's photosphere were of uniform brightness, then the fractional decrease in apparent brightness of the star is the area ratio ($\pi r^2/\pi R^2$), where $\pi r^2$ is the projected area of the planet, radius $r$, and $\pi R^2$ that of its star, radius $R$. In fact the photosphere of a star appears to dim slightly towards the edge (the limb). This limb darkening arises because the radiation we receive from the limb is predominantly from the uppermost regions of the photosphere, which are cooler than deeper down. This modifies the light curve, depending on which chord of the star's disc the planet traverses. With a sufficiently precise light curve a correction can be applied. Also, dips in apparent brightness can arise from events other than a planetary transit, such as a grazing transit by a fainter companion star. This, and other types of extraneous events, can be identified from the shape of the light curve.

Figure 4    The apparent decrease in the observed brightness of a star when one of its planets passes between us and its star. The fractional decrease does not depend on the distance from us to the star.

If, far beyond the Solar System, a Jupiter-sized planet were to be observed transiting the a star like the Sun, then the dip in apparent stellar brightness would be about 1% (Jupiter is about a tenth of the radius of the Sun). Ground based photometric precision is limited by fluctuations in the transparency of the atmosphere to about 0.01%, so Jupiter's light curve could be measured with considerable accuracy. However, an Earth-sized planet (a hundredth of the Sun's radius) would produce a dip of around 0.01 %, so would not be detectable. Only for the abundant M dwarfs – stars with radii the order of a tenth that of the Sun – would Earth-sized planets be detectable from the ground[4]. Things are better in space, where the limit is imposed by variations in the star's brightness. These can be of the order of the transit time, a few hours, but give a different shaped light curve, that does not repeat regularly. In the most favourable cases, a planet about a third of the Earth's radius, orbiting an M dwarf and observed over several transits, could be detected in a short period orbit (a few tenths of an AU).

There are many ground-based searches for transits. In space the Hubble Space Telescope has been used. The French-led CoRoT space telescope includes transit searches in its programme, and has been gathering data since 02 February 2007 (COROT 2008). NASA's Kepler Space Observatory, dedicated to transit searches, is scheduled for launch in April 2009, and will be the first instrument capable of detecting Earth-sized planets and smaller (Kepler 2008).

---

[4] The Sun is classified as a G dwarf. Dwarfs stars are in the longest phase of their lifetime, the main sequence phase, converting hydrogen to helium in their cores. In decreasing mass, dwarfs are labelled O, B, A, F, G, K, M. As mass decreases the luminosity and radius decrease, but the abundance of the stars increases.

Figure 5 shows some transit data. The interval between the transits gives the orbital period, and when the mass of the star can be estimated (which is often the case), we get the semimajor axis of the planet's orbit.

> Figure 5    The light curve of the F dwarf WASP-12 due to the transit of a planet with a radius 1.9 times that of Jupiter, and in a very small orbit, period 1.09 days. The light curve is shown only in the vicinity of the transit. This planet was discovered with the 2.0 metre aperture Liverpool Telescope high on La Palma in the Canary Islands. (SuperWASP data, private communication from Leslie Hebb)

The probability of transit is approximately $R/a$, where $R$ is the radius of the star and $a$ is the semimajor axis of the planet's orbit. Given that the orbits of exoplanetary systems have random values of inclination $i$ (Figure 2), approximately 10% of solar radius stars with a planet in a 0.05 AU orbit will transit their star[5]. At about 5 AU (Jupiter's distance from the Sun), the proportion falls to about 0.1%.

A few tens of exoplanets have been discovered through their transits, making it second only to the radial velocity (RV) method in numbers of discoveries. Many of these exoplanets have also been detected by radial velocity measurements. In a few cases such measurements preceded the transit observations, but in most cases they were subsequent to them. In all cases we obtain the radius $r$ of the planet. We also have its actual mass $m$ – to be observed in transit the value of the inclination $i$ must be close to 90°. We thus have the mean density of the planet, $m/[(4\pi/3)\,r^3]$, which provides a powerful constraint on its composition.

In some cases we can learn more about the planet. We receive lots of infrared radiation from the star plus a much smaller amount emitted by its planet. When a planet passes behind its star the infrared radiation we receive thus falls slightly, and this has lead in a few cases to an estimate of the temperature of the source of the planet's infrared radiation, which is its atmosphere, or surface, or some combination (e.g. Deming et al. 2006). In a few other cases it has been possible to detect the radiation that identifies specific elements in the planet's atmosphere, such as sodium (Snellen et al.

---

[5] The semimajor axis of the Earth's orbit is one astronomical unit, 1 AU. This is 149.6 million kilometres.

2008).

Earth-mass planets not directly detectable by transit could be detected by accurate timing of the transits of much larger planets. If accuracies of a few seconds can be achieved then any periodic variations in the interval between transits would be due to the gravitational tugging of another planet, whose mass could be determined if the mass of the larger planet were known. This technique has been used on two targets, each having one known planet, with ground based photometry combined with Hipparcos photometry. No extra planets have yet been discovered.

*Gravitational microlensing*

This method relies on the lens effect of a foreground star on the light we receive from a background star. When the alignment is exact you might think that the background star would be hidden from view. Instead, the gravitational field of the foreground star bends towards us the light from the background star that passes near it. The foreground star has thus acted as a gravitational lens, and is called the lensing star. With sufficient spatial resolution we would see the background star as a ring, called the Einstein ring after Albert Einstein, whose theory of General Relativity explains the bending in detail. The Einstein ring has a radius (in radians) given by

$$\theta = [(4GM/c^2)(D-d)/(D\,d)]^{1/2} \qquad (6)$$

where $D$ is the distance to the background star, $d$ is the distance to the lensing star, and $c$ is the speed of light. For typical values $\theta$ is of the order of a few hundred micro-arcseconds. This is too small for the ring to be resolved with currently available telescopes, but the lensing star directs more light from the background star towards the observer than is received from the unlensed background star. Thus, as the motions through space of the background and lensing stars produce the alignment, an apparent brightening of the background star is observed.

Alignments are never exact, and what we would observe with sufficient spatial resolution is shown in Figure 6(a) for a specific case. You can see that at any instant the background star is imaged in the form of two smudges, …… (1, 1') then (2, 2'), then (3, 3'), and so on. The wide smooth curve in Figure 6(b) shows the corresponding variation in apparent brightness of the background star. The amplification $A$ reaches a peak of about 5, corresponding to a five times apparent brightening of the background star. The value of $A$ depends on the two stars' angular separation on the plane of the sky as a multiple $f$ of $\theta$

$$A = (f^2 + 2)/[f(f^2 + 4)^{1/2}] \qquad (7)$$

You can see that this is in accord with the light curve in Figure 6(b): at separations well beyond $\theta$, $f \gg 1$ and $A \approx 1$. $A$ then increase as $f$ decreases, up to the closest approach, whereafter $f$ increases and $A$ decreases. The encounter lasts for a duration of order

$$t = \theta/n \qquad (8)$$

where $n$ is the rate at which the two stars' angular separation on the plane of the sky changes. Typically, $t$ is a few days.

> Figure 6  (a) Gravitationally lensed images of a background star by a foreground (lensing) star (the smudges labelled 1, 1', 2, 2', 3, 3') with a planet at ×. (b) The light curve corresponding to the encounter in Figure 6(a).

What if a planet is present? If it is present at the location marked × in Figure 6(a) then the upper lensed image of the background star will sweep across it. The gravitational field of the planet then produces the spikes in the light curve in Figure 6(b). Figure 6(b) shows that the planet's lensing event is comparatively brief, easily missed, and therefore the practical approach is to use the start of the *stellar* lensing as an alert. The planet's peaks are taller the closer the planet is to the background star's image, independent of the mass of the planet. The duration of planet's lensing, $t_p$, is of order $\theta_p/n$ where $\theta_p$ is the angular radius of the planet's Einstein ring (equation 6 with $m$ in place of $M$). It thus follows that

$$m = M (t_p/t)^2 \qquad (9)$$

where $m$ is the mass of the planet. So, $m$ can be readily obtained from the light curve, even small values of $m$. But all we learn about its orbit is the planet's projected distance on the sky during the lensing.

By 15 September 2008, eight planets had been discovered by microlensing, two orbiting one star. Most discoveries have been made by the OGLE and MOA surveys (OGLE 2008, MOA 2008. There is also the PLANET + RoboNet survey (RoboNet 2008). Note that such surveys also find planets through transit photometry – both are photometric techniques.

*What we can learn about exoplanets from each detection method*

Table 1 summarizes what we can learn about exoplanets from each detection method.

Table 1    Planet properties obtained (1)

|  | Direct imaging | Transit photometry | Astrometry | Radial velocity measurements | Gravitational microlensing |
|---|---|---|---|---|---|
| Mass | Yes (2) | No | Yes | Min. value (3) | Yes |
| Radius | Yes | Yes | No | No | No |
| Semimajor axis | Yes | From period | Yes | From period | Projected value |
| Orbital period | Yes | Yes | Yes | Yes | No |
| Orbital eccentricity | Yes | No | Yes | Yes | No |

1 Some properties of the star are needed, notably the mass, and, for transit photometry, the radius
2 From the orbit of the star
3 In some cases the actual mass can be obtained

As pointed our earlier, from transit photometry, which gives us the radius $r$ of the planet, and the planet's mass $m$ from any other method (which will be the actual value, the planet having been observed in transit), we get the mean density of the planet, $m/[(4\pi/3)\,r^3]$, which provides a powerful constraint on its composition.

**The known exoplanets**

The first (non-pulsar[6]) exoplanet was discovered in 1995, in orbit around the solar-type star 51 Pegasi (Mayor & Queloz 1995). The planet is nearly half the mass of Jupiter, and has a circular orbit with a semimajor axis of only 0.052 AU (Mercury's is 0.387 AU). As of 15 September 2008, 305 non-pulsar planets had been confirmed in 263 (non-pulsar) systems. There are 29 multiple planet systems. The great majority have been discovered through radial velocity measurements. Several tens have been discovered by transit photometry, and fewer than ten by all other detection methods combined. A few tens of planets have been detected by more than one method, notably those detected by radial velocity measurements and transit photometry. As explained above, this enables the mean density of the planet to be calculated. These data have been obtained from a catalogue of exoplanetary systems that can be found at http://www.exoplanet.eu/ (Schneider), which is updated on a roughly weekly basis. It also contains much relevant information and links to research papers.

Table 2 summarizes the properties of the currently known (non-pulsar) exoplanetary systems.

---

[6] A pulsar is a remnant of a massive star that exploded. Two pulsars are known to have planets, one with three planets, the other with a single planet. It is highly unlikely that pulsar planets could be inhabited.

Table 2  A summary of some properties of the known (non-pulsar) exoplanetary systems

| Characteristic (1) | Data | Comment |
| --- | --- | --- |
| Stellar mass/solar mass (2) | 0.34-1.5 | Only one star is not a dwarf (3). M dwarfs are increasingly targeted. |
| Stellar distance/light years | ≥10.5 | Very few beyond ≈ 300 light years – mostly from microlensing |
| Planet mass/mass of Jupiter | 0.01-13 | 0.01 is 3.2 Earth masses; ≥13 likely to be brown dwarf stars |
| Planet orbit semimajor axis/AU | 0.0177-5.77 | The four detected by imaging have orbits ranging from 46-670 AU |
| Planet orbital eccentricity | 0-0.92 | Most planets in small orbits have small orbital eccentricities |

(1) The slash denotes the units e.g. stellar masses are in units of the Sun's mass.
(2) Five of the stars are in triple star systems, and about a dozen are in binary systems. In every case the planet(s) orbits just one of the stars.
(3) The exception is a subdwarf star with a planet inferred from modulation of its pulsations.

*The stars*

You can see from Table 2 that the stars in the known (non pulsar) exoplanetary systems are dwarfs i.e. stars in the main sequence phase of their lifetime (footnote 4). This is an observational selection effect. Planets can be consumed or ejected after the main sequence phase as the star swells enormously. Also, the post main sequence phases are much shorter and much more violent than the main sequence phase – not at all conducive to planetary survival and certainly not to habitability (Green and Jones 2004, chapters 6 & 7). It is thus not surprising that dwarfs have been the focus of our searches.

Not all dwarfs have been equally treated, with searches favouring F, G, and K dwarfs (footnote 2). There are several reasons for this. The main ones are as follows.
- They are fairly luminous, account for about 20% of dwarf stars, have good surface stability, and plenty of sharp spectral lines for the radial velocity method to work on.
- They have main sequence lifetimes longer than about 3000 million years, long enough for life to emerge on suitable planets and become detectable from afar, if the Earth is any guide (though perhaps 10% of them are not yet this old). The massive O, B, and A dwarfs have shorter main sequence lifetimes, and in any case constitute only about 1% of dwarf stars.

The M dwarfs are now being targeted. They have main sequence lifetimes longer than the 13 600 million year age of the Universe, and comprise nearly 80% of dwarfs. They are intrinsically faint, making observations difficult, which is one reason why they were earlier ignored.

Another consideration is the so-called metallicity of the star. This is its proportion of elements heavier than hydrogen and helium, elements that are essential for the formation of all planets. The Sun's metallicity is 1.6%, and nearly all of the stars within 300 light years, the range within which most exoplanets have been detected (Table 2), have values exceeding 0.5%, probably sufficient for planet formation (Pudritz et al. 2007).

*Minimum masses*

Figure 7 shows, for the known exoplanets, the (minimum) mass versus semimajor axis. In some cases, such as when the planet has also been observed in transit, the value of *i* (equation (5)) is close to 90°, and so this will be the actual mass. In other cases, if *i* is assumed to be randomly distributed, then, on average, the actual mass is 1.3 times the minimum mass, and the great majority will have an actual mass less than twice the minimum. Therefore, if actual masses were available for all exoplanets, the distribution in Figure 7 would move upwards, but only slightly on the logarithmic scale in the Figure.

Figure 7    The minimum mass versus orbital semimajor axis of the known exoplanets. Jupiter, Saturn, and the Earth have been added for comparison, and the semimajor axis of Mercury is indicated. Mercury has a mass off the bottom of the diagram, at 0.000174 Jupiter masses.

The smallest mass so far recorded (as of 15 September 2008) is just 3.2 Earth masses (3.2 $m_E$), though with a large uncertainty such that the two sigma range is 1.7-8.2 $m_E$. The discovery was made by gravitational microlensing (Bennett et al. 2008) and so the mass is the actual value. The distance from its star, about 0.6 AU, is the projected value, not the actual semimajor axis. The star, MOA-2007-BLG-192-L, has a mass only about 6% that of the Sun. It is roughly 3000 light years away.

The upper end of the exoplanet mass range should be around 13 Jupiter masses (13 $m_J$). This is because objects with greater masses, up to about 80 $m_J$, are more like stars than planets in their interior structure, and are called brown dwarfs. They are fluid and convective throughout, and unlike planets lack a layered composition. They are certainly not habitable. They have no main sequence phase – only at masses greater than about 80 $m_J$ is there such a phase, the M dwarfs being the lowest mass main sequence stars. In Figure 7 you can see that minimum masses up to about 20 $m_J$ have been included. This is because the 13 $m_J$ threshold is uncertain. On the other hand some minimum masses safely below 13 $m_J$ could correspond to actual masses of what are undoubtedly brown dwarfs.

The blank area in Figure 7 corresponds to where it is more difficult to detect exoplanets – planets in large orbits with long orbital periods, and planets in not-so-large orbits with low minimum mass. It is expected that this zone will fill in as instruments and techniques improve, and as data are accumulated. Therefore, more planets with masses less than about 0.03 $m_J$ (about 10 $m_E$) will be discovered. You will see later that these are the sort of planets most likely to be inhabited. You will also see that they would have to orbit at the right distance from the star for water to be stable as a liquid on and near the planet's surface i.e. the orbit has to lie in the classical habitable zone (HZ). The extent of this zone depends (mainly) on the mass of the star, the less massive the lower the luminosity, and the closer the zone to the star. Overall, this zone for dwarfs spans the approximate range 0.1 AU to about 20 AU (Jones et al. 2006). It is thus the strip along the bottom centre of Figure 7 where habitable planets will be found.

*Hot Jupiters*

You can see in Figure 7 that many planets are in very small orbits, smaller than Mercury's orbit, the planet nearest the Sun. These constitute the "hot Jupiters" – certainly hot, with atmospheric temperatures of order 1000 K, and certainly with masses of the order of $m_J$. It is the high gravity of hot Jupiters that prevents the thermal escape of their atmospheres. But in the Solar System, the giant planets Jupiter and Saturn are far from the Sun, as Figure 7 shows. So, why are there hot Jupiters?

"Hot Jupiters" could not have formed where we find them today. Our understanding of the formation of planetary systems leads firmly to the conclusion that they must have formed further out, where there was a greater abundance of water, which, along with rocky materials, formed huge kernels that could capture the nebular gases, notably hydrogen and helium (Jones 2007 Chapter 2). Therefore, the hot Jupiters must have started forming further out and migrated inwards as they continued to grow. This is thought to occur through the gravitational interaction of the growing giant planet with the circumstellar disc of gas and dust from which it is being born. Migration needs to be halted if the planet is not going to end up in the star – there are various plausible mechanisms. See (Jones 2004 p223) for further details.

*Composition of exoplanets*

The best data we have from which the composition of an exoplanet can be inferred, is the mean density of the planet. Jupiter, has a mean density of 1330 kg m$^{-3}$, whereas that of the Earth is 5520 kg m$^{-3}$. The difference would be even larger without greater self compression in Jupiter than in the 318 times less massive Earth. We know, from its low density and a host of other measurements, that Jupiter is dominated by the intrinsically least dense elements, hydrogen and helium, layered, and perhaps with its rocky and icy materials (mainly water) concentrated into a core (Irwin 2006, section 2.7.3). It has a hot interior, and is fluid throughout. Saturn, 95 $m_E$, is broadly similar, but with a greater proportion of hot rocky and icy materials (Irwin 2006, section 2.7.3). By contrast, the Earth is dominated by the intrinsically more dense rocky materials, notably silicates and iron, and, except for the outer part of its iron core, is solid.

Nearly 30 exoplanets have had their densities determined through a combination of radial velocity measurements and transit photometry. The masses cover the approximate range 0.33-1.3 $m_J$, and the mean densities 300-1500 kg m$^{-3}$, with little correlation between the two. None is far from their star, so each is likely to be bloated, in part due to its hot atmosphere, and in part because its interior would not have cooled as much as it would have had it not migrated inwards. All are surely dominated by hydrogen and helium.

For all the other exoplanets we can only make reasonable assumptions. These result in the following conclusions about their compositions. (See Jones 2008 p169 for further details.)
- Above roughly 50 $m_E$: broadly like Jupiter and Saturn.
- In the approximate range 10-50 $m_E$: broadly like Uranus (14.5 $m_E$) and Neptune (17.1 $m_E$), which are far less dominated by hydrogen and helium, consisting largely of hot icy and rocky materials (Irwin 2006, section 2.7.4).
- Below about 10 $m_E$: dominated by rocky materials (like the Earth), but perhaps water-rich at the upper end of the mass range, unless the planet is very close to its star.

It is this last category that have surface and near-surface regions suitable for life, provided that water in these regions can be stable as a liquid – see below. As yet, very few rocky planets have been discovered in this astrobiologically relevant range.

In a few cases, radiation has been detected from an exoplanet. Atmospheric temperatures for a few hot Jupiters has been determined by comparing the infrared radiation from the star plus planet when it was in transit, with that when it was behind the star. Temperatures of order 1000-2000 K have been detected for the region in the planet's atmosphere from which the infrared radiation to space is emitted (e.g. Charbonneau et al. 2005, Deeming et al. 2005, 2006). Additionally, a small number of exoplanets have had some minor atmospheric constituents detected in their infrared spectra, for example, sodium vapour (Snelling et al. 2008), silicate dust (Richardson et al. 2007), water vapour (Barman 2008), and methane (Swain et al. 2008). Note that hydrogen and helium have weak infrared spectra, so it is unsurprising that these gases have not yet been detected within the atmosphere, though for one planet hydrogen has been detected high up in a hugely inflated atmosphere (e.g. Ben-Jaffel, 2007).

*Orbital eccentricities*

Figure 8 shows, for the known exoplanets, the orbital eccentricity versus semimajor axis. You can see that whereas small eccentricities, such as those of Jupiter and Saturn, are found over the full range of orbital semimajor axes, the high eccentricities are increasingly found at larger semimajor axes. The lack of high eccentricities at small semimajor axes is due to tidal interactions with the star, one effect of which is to reduce eccentricity. But why are some of the eccentricities at larger semimajor axes so much greater than those found in the Solar System? The formation of planets does not lead to this. Two possible answers each require the presence of two giant planets. In one case the orbital periods need to be in a simple numerical ratio to produce what is called a mean motion resonance. The other case requires a close encounter between two giants, a common outcome being the ejection of one of them and the acquisition by the other of a high eccentricity orbit. See Jones (2004 pp223-224) for further details.

>Figure 8    The orbital eccentricity versus orbital semimajor axis of the known exoplanets. Jupiter and Saturn are shown for comparison.

**The unknown exoplanets**

Table 2, and Figures 7 and 8 summarize many of the attributes of the known exoplanets. Additionally, it has been estimated that at least 25% of the F, G, and K dwarfs within a few hundred light years of the Sun have one or more planets (Lineweaver & Grether 2003). As yet the proportion known is nearer to 10%, but it can only increase as the more difficult detections are made, of lower mass planets, and planets further from their star. Also, many of the systems known to have just one planet, are likely to have more, the less easily detectable planets that could be detected by further investigations. We will also increase the distance range at which meaningful statistics can be obtained.

Figure 9 shows where developments in instruments and techniques will take us in the next few decades. We will surely discover habitable planets, perhaps in abundance.

Figure 9   Where, in the discovery of exoplanets, developments in instruments and techniques will take us in the next few decades. Note that the arrows show trends, and do not imply convergence at a particular orbital period.

**Habitable exoplanets**

At present there is just one discovery of an exoplanet that could be habitable (Udry et al. 2007). This is Gliese 581c, the innermost of the three planets of the M dwarf Gliese 581, 20.5 light years away. It has a minimum mass of 5.03 $m_E$ (from radial velocity measurements), and though it is only 0.073 AU from its star, Gliese 581 is so faint that the planet orbits in the HZ.

*The HZ*

It is in the HZ that we expect to find habitable planets. The HZ is the range of distances from the star within which radiation from the planet's star would maintain water in liquid form at the surface of a rocky planet with a significant atmosphere. At closer distances all surface water would have evaporated, and at further distances it would be frozen. The HZ depends on the model adopted for the test planet's atmosphere, notably on its mass and composition, which will determine the size of the greenhouse effect, and on the cloud cover, which will determine how much of its star's radiation is reflected away. A widely accepted model is that of Kasting et al. (1993). Their model contains several criteria for defining the HZ and one of these yields a HZ for the Solar System shown in Figure 10. In fact, two HZs are shown. The shaded annulus is for the Sun today; the dashed circles for the Sun at its birth, 4600 million years ago when its luminosity was only about 70% its present value. It's "surface" temperature was also less then, which adds slightly to the luminosity effect (Jones 2004 p83). As the luminosity and surface temperature grew the HZ moved outwards.

Figure 10   The HZ in the Solar System today (shaded annulus) and at the Sun's birth (the dashed circles). The solid lines show the orbits of the four terrestrial planets.

If the HZ makes sense in the Solar System then the model can be used to determine the HZ in exoplanetary systems, by substituting the luminosity and surface temperature of the star. The HZ today includes the Earth, which is reassuring! It does not include Venus – the model predicts that Venus is too hot for liquid water at its surface, and this is in accord with its mean global surface temperature of 737 K (Williams 2005). Mars is the instructive case. Figure 10 shows it to be just within today's HZ, and yet liquid water cannot be stable anywhere on the surface of Mars, except for a few hours now and then at low altitudes. This is because the atmosphere is so thin that the surface temperature on Mars only exceeds 273 K for a few hours at low latitudes, low altitudes, and at midday – the surface of Mars almost everywhere, almost all the time, is well below 273 K (Williams 2007). The thin atmosphere – 636 Pa pressure at mean radius (variable from 400 to 870 Pa depending on season[7]) – is due to the low mass of Mars, 0.107 $m_E$. The associated low gravity has resulted in some of its atmosphere being lost to space. The low mass has also led to insufficient geological activity that would otherwise have recycled atmospheric gases from where they had been incorporated in surface and subsurface reservoirs.

It is estimated that a mass of at least 0.3 $m_E$ is necessary for a substantial atmosphere (Williams et al. 1997, Raymond et al. 2007). The upper limit on mass, before thick atmospheres of hydrogen and helium are likely to be present, is of order 10 $m_E$.

*The stability of exo planets in the HZs*

In order for life to be present today on the surface of an exoplanet it must have been in its star's HZ for at least the past few thousand million years for life to have affected its atmosphere to an extent that could be detected from afar (if the Earth's biosphere is any guide – see Figure 11 below and the associated discussion). If a giant planet is too close to the HZ, its gravitational field would remove planets from the HZ. A study has shown that roughly half of the known exoplanetary systems have HZs that have such prolonged stability, in at least part of the HZ (Jones et al. 2006). Provide that suitable planets formed, it follows that about half of the known exoplanetary systems could have detectable biospheres.

You have seen that to create hot Jupiters it is thought that migration has occurred. Given that they are hot, this must necessarily involve migration through the HZ. At such a time there would have been no planets in the HZ, only embryos and planetesimals. These would have been scattered. So does such migration preclude the formation of planets in the HZ? If it does, this reduces to less than 10% the proportion of the known exoplanetary systems that could have detectable biospheres (Jones et al. 2006). Fortunately, studies have shown that the post-migration formation of suitable planets is not ruled out by such migration. Fewer planets might form in the HZ, but at least one suitable planet is a likely outcome (Fogg and Nelson 2007).

---

[7] The triple pressure of water is 610 Pa, so the low temperature is the main problem.

*The issue of bombardment*

As well as orbital stability, for a planet to develop a biosphere it must not suffer bombardments so frequently that the biosphere clock is often reset. It has been argued that a giant planet beyond the HZ, of the order of Jupiter's mass, is necessary to provide sufficient shielding. However, recent work has cast doubt on this – the issue is not clear cut (Horner & Jones 2008).

*Beyond the HZ*

It must be emphasised that the HZ is predicated on the star heating a planet's surface. A planet can be habitable below its surface, such as the deep hot biosphere on Earth (Gold & Dyson 1998).

Also, a large satellite can be tidally heated by its planet, again to produce a subsurface biosphere, such as might be the case in the oceans beneath the icy carapace of Jupiter's Europa. Europa, along with the other three large satellites of Jupiter (Io, Ganymede, and Callisto), are large. Io and Europa are roughly the size of the Moon, and Ganymede and Callisto are roughly the size of Mercury. They are thought to have formed during the formation of Jupiter, from a disc of debris encircling the planet (Jones 2007 Section 2.3.1). There is no reason to expect things to have been different in exoplanetary systems with giant planets. Such potential habitats could be widespread in the cosmos.

Jupiter's large satellites are tightly bound to the planet, and are likely to have been able to remain bound throughout the gentle process of migration. If a giant planet with a satellite with a mass around 0.3 $m_E$ (or larger), ended in the HZ it would not need tidal heating to be habitable. Such potential habitats could also be widespread in the cosmos.

However, if we are to detect life from afar, then surface life offers the best prospect, and that is likely to be confined to sufficiently massive planets (or planetary satellites) in the HZ.

**How to tell from afar whether a planet is habitable/inhabited**

The best way of determining from afar whether an exoplanet is habitable/inhabited is from the planet's infrared emission spectrum. Other techniques can be found in Jones (2008, Section 13.6).

Figure 11 shows the infrared emission spectrum of the Earth. It is representative of the spectrum that an alien would obtain from afar, with a telescope plus spectrometer with exquisite spectral resolution. The spectrum is the result of black body radiation emitted by the Earth's warm surface. After this radiation has traversed the Earth's atmosphere, certain wavelengths have been attenuated. How can we tell from this spectrum that the Earth is very probably inhabited?

> Figure 11   The infrared emission spectrum of the Earth, obtained in daytime by the Nimbus 4 satellite in the 1970s, over a cloud-free part of the western Pacific Ocean. Adapted from Figure 8.5 in Fundamentals of Atmospheric Physics, by M L Salby, Academic Press, 1995.

The first indication is the fit of the 300K black body curve to the right hand part of the spectrum. This indicates that the Earth's surface temperature is about 300 K, certainly warm enough for liquid water, if the pressure is sufficient. The pressure broadening of some of the spectral absorption lines indicates that it is. That water is present is indicated by the fine detail in the spectral regions indicated. These details are from water vapour in the atmosphere. The fit at 275 K at left is from the atmosphere at altitude – in the troposphere the temperature generally declines as altitude increases.

The large absorption due to $CO_2$ is clear. Though $CO_2$ accounts for only a fraction $0.37 \times 10^{-3}$ of the molecules in the Earth's atmosphere, it has very strong absorptions in the infrared, so much so that the radiation at these wavelengths that makes it into space originates from high in the atmosphere where the temperatures are around 225 K (Figure 11). Thus there is carbon on Earth, no great surprise given that, after hydrogen and helium, carbon is second only to oxygen in the abundance of the chemical elements in the Universe.

So, the Earth is habitable by carbon-liquid water life. But where's the indication in Figure 11 that it is actually inhabited? One such is the ozone ($O_3$) absorption feature. In the Earth's atmosphere $O_3$ is derived from $O_2$ in the upper atmosphere through the action of solar UV radiation. $O_2$ has a very weak infrared absorption spectrum, whereas that of $O_3$ is very strong. Thus, $O_3$ acts as a proxy for $O_2$. The strength of the $O_3$ feature in Figure 11 indicates that there is a copious source to maintain the $O_2$, which otherwise would vanish in the oxidation of surface rocks and reduced volcanic gases. We know that it is oxygenic photosynthesis that sustains the $O_2$ (and even gave rise to it). Therefore the $O_3$ feature indicates the presence of a biosphere.

But $O_3$ can also result from the UV photodissociation of water vapour high in the atmosphere, and though today the upper atmosphere is too dry for this mechanism to contribute much, there can be short periods when the atmosphere is much damper, such as through high volcanic activity. The clincher is the methane ($CH_4$) feature in Figure 11. Though it accounts for only about a fraction $2 \times 10^{-6}$ of the molecules in the (lower) Earth's atmosphere, it has very strong absorptions in the infrared. The surprise is, that in an $O_2$ rich atmosphere there is anywhere near this much $CH_4$. Methane is generated by certain bacteria (including some of those in the guts of ruminants), in marshes (methane is also known as marsh gas), and in paddy fields. Without a substantial rate of generation by the biosphere, the amount of $CH_4$ would decline hugely in a matter of decades. Thus it is the presence of $O_2$ and $CH_4$ together, far from chemical equilibrium, that provides very strong evidence for a biosphere.

When we have the instrumental capability to see an exoplanet, even as a single pixel, Figure 12 shows the sort of infrared emission spectrum that we might obtain of an Earth-type planet in the HZ of a solar-type star. In this example the star is about 30 light years away, the instrument is the proposed infrared space telescope Darwin (ESA 2006), and the exposure time is 40 days. The spectral resolution is far inferior to that in Figure 11, so much so that the $CH_4$ feature is not seen. We could conclude that this exoplanet is habitable, perhaps inhabited, but we lack the clincher that it is inhabited. Alas! this is a likely outcome – very frustrating.

> Figure 12    The infrared emission spectrum that might be obtained by a Darwin sized infrared space telescope, with a 40 day exposure of an Earth-type planet in the HZ of a solar-type star about 30 light years away.

But what if the $O_3$ feature were absent? We would then have no indication at all that the exoplanet is inhabited. But absence of evidence is not evidence of absence. There are several possibilities.
- Plenty of $O_2$ but too little stellar UV to form much $O_3$
- Efficient removal of $O_3$
- Anoxygenic photosynthesis, which some microbes on Earth perform
- No photosythesis, only chemosynthesis, which some microbes on Earth perform
- A planet too young for oxygenic photosynthesis to have caused observable effects on the atmosphere ($O_2$ in Earth's atmosphere only rose substantially after about 2400 million years ago, Canfield 2005)
- A biosphere beneath the planet's surface, such as the hot, deep biosphere on Earth (Gold &

Dyson 1998).

We thus might have to be satisfied with "habitable, perhaps inhabited" as far as carbon-liquid water life is concerned. However, were we to find spectral signatures of a pair of gases other than $O_3$ and $CH_4$, far from chemical equilibrium, it would not be unreasonable to speculate than a biosphere with a truly alien biochemistry was present.

Detection methods are advancing rapidly, and data are being acquired apace. Our knowledge of exoplanetary systems is building, though it will be at least a decade before space telescopes like Darwin can provide the spectra required to investigate the habitability/inhabitation of an exoplanet.

To be absolutely sure that an exoplanet is inhabited we would have to send a probe there, but that is a distant prospect even for the nearest exoplanetary systems. Of course, intelligent life could tell us it was out there, by sending us signals. But that's another story.

**Conclusions**

Many exoplanets have been discovered, and the trickle of discoveries of planets in the HZ will surely increase to a steady flow. Among these will be rocky bodies (perhaps with deep oceans) with masses greater than about 0.3 Earth masses, the most likely planets to bear surface life. Surface life can be detected from afar provided that we can obtain the spectrum of the planet – the infrared emission spectrum offers the best prospects. ESA's Darwin mission, should, in a decade or so, give us this capability. We are thus on the threshold of the possibility of discovering that we are not alone.

**Acknowledgements**

I thank Mark Burchell for the invitation to present a review at the 3$^{rd}$ conference of the Astrobiology Society of Britain, on which this paper is based, and to Leslie Hebb for Figure 5.

**References**

Barman, T.S. (2008). On the Presence of Water and Global Circulation in the Transiting Planet HD 189733b. *Astrophys. J. Letters*, **676**, L61–L64.
Ben-Jaffel, L. (2007). Exoplanet HD209458b: Inflated hydrogen atmosphere but no sign of evaporation. *Astrophys. J.* **671**, L61-64.

Bennett, D.P. et al. (2008). A low-mass planet with a possible sub-stellar–mass host in microlensing event MOA-2007-BLG-192. *Astrophys. J.*, 684, 663-683.

Canfield, D.E. (2005). The early history of atmospheric oxygen: homage to Robert M. Garrels. *Ann.Rev. Earth Planet. Sci*. **33**, 1-36.

Charbonneau, D. et al. (2005). Detection of thermal emission from an extrasolar planet. *Astrophys. J.* **626**, 523-529.

CoRoT (2008). CoRot overview. http://smsc.cnes.fr/COROT (accessed September 2008)

Deming, D. et al. (2005). Infrared emission from an extrasolar planet. *Nature*. **434**, 740-743.

Deming, D. et al. (2006). Strong infrared emission from the extrasolar planet HD189733b. *Astrophys. J.* **644**, 560-564.

ESA (2006). Darwin overview. http://www.esa.int/esaSC/120382_index_0_m.html (accessed August 2008)

Fogg, M.J & Nelson, R.P. (2007). On the formation of terrestrial planets in hot-Jupiter systems. *Astron. & Astrophys*. **461**, 1195-1208.

Gaia (2008), Gaia overview. http://www.rssd.esa.int/index.php?project=GAIA&page=index (accessed September 2008)

Gold, T. & Dyson, F.J (1998). The deep, hot biosphere. *Springer Heidelberg and New York*.

Green, S.F. & Jones, M.H. (editors) (2004). An Introduction to the Sun and Stars. *Cambridge University Press*.

Hipparcos (2007). Hipparcos overview. http://www.rssd.esa.int/index.php?project=HIPPARCOS&page=fposter (accessed September 2008)

Horner, J.A. & Jones, B.W. (2008). Jupiter – friend or foe? I: the asteroids. *Inter. J. Astrobiol*, **7**, 251-261.

Irwin, P. (2006). Giant planets of our Solar System: an introduction. *Springer-Praxis*, Heidelberg/Chichester UK

Jones, B.W. (2004). Life in the Solar System and Beyond. *Springer-Praxis*, Heidelberg-Chichester.

Jones, B.W., Sleep, P.N. & Underwood D.R. (2006). Habitability of known exoplanetary systems based on measured stellar properties. *Astrophys. J.* **649**, 1010-1019.

Jones B.W. (2007). Discovering the Solar System 2$^{nd}$ edition. *John Wiley & Sons*, Chichester UK.

Jones, B.W. (2008). The Search for Life Continued – Planets around Other Stars. *Springer-Praxis*, Heidelberg & New York.

Kasting, J.F., Whitmire, D.P. & Reynolds, R.T. (1993). Habitable zones around main sequence stars. *Icarus* **101**, 108-128.

Kepler (2008). Kepler Space Observatory overview. http://kepler.nasa.gov (accessed September

2008)

Lineweaver, C.H. & Grether, D. (2003). What fraction of Sun-like stars have planets? *Astrophys. J.* **598**, 1350-1360.

Mayor, M. & Queloz, D. (1995). A Jupiter-mass companion to a solar-type star. *Nature* **378**, 355-359.

MOA (2008). MOA overview. http://www.phys.canterbury.ac.nz/moa/index.html (accessed September 2008)

OGLE (2008). OGLE overview. http://ogle.astrouw.edu.pl (accessed September 2008)

Perryman, M.A.C. (2000). Extra-solar planets. *Rep. Prog. Phys*. **63**, 1209-1272.

PLANET+RoboNet (2008). Overview of project. http://www.astro.ljmu.ac.uk/RoboNet (accessed September 2008)

PRIMA (2007). Overview of PRIMA at the VLT. http://www.eso.org/projects/vlti/instru/prima/index_prima.html accessed September 2008)

Pudritz, R., Higgs, P. & Stone, J. (editors) (2007). Planetary Systems and the Origin of Life. *Cambridge University Press*.

Raymond, S.N., Scalo, J. & Meadows, V.S. (2007). A decreased probability of habitable planet formation around low-mass stars. *Astrophys. J*. **669**, 606-614.

Richardson, L.J. et al. (2007). A spectrum of an extrasolar planet. *Nature* **445**, 892-895.

Salby, M.L. (1995). Fundamentals of Atmospheric Physics. *Academic Press*, San Diego.

Schneider, J. The extrasolar planets encyclopaedia. http://www.exoplanet.eu/ (accessed September 2008)

SIM PlanetQuest (2008). Overview of project. http://planetquest.jpl.nasa.gov/SIMLite/sim_index.cfm (accessed September 2008)

Snellen, I.A.G., Albrecht, S., de Mooij, E.J.W. & Le Poole, R.S. (2008). Ground-based detection of sodium in the transmission spectrum of exoplanet HD 209458b. *Astron & Astrophys* **487**, 357-362.

Swain, M.R., Vasisht, G. & Tinetti, G. (2008). The presence of methane in the atmosphere of an extrasolar planet. *Nature* **452**, 329-331.

Udry, S. et al. (2007). The HARPS search for southern extra-solar planets. *Astron & Astrophys.* **469**, L43-L47.

Vogt, S.S., Butler, R.P., Marcy, G.W., Fischer, D.A., Henry, G.W., Laughlin, G., Wright, J.T., & Johnson, J.A. (2005). Five new multicomponent planetary systems. *Astrophys. J.* **632**, 638-658.

William, D.M., Kasting, J.F. & Wade, R.A. (1997). Habitable moons around extrasolar giant planets. *Nature* **385**, 234-236.

Williams, D.R. (2005). NASA fact sheet,

http://nssdc.gsfc.nasa.gov/planetary/factsheet/venusfact.html (accessed August 2008)

Williams, D.R. (2007). NASA fact sheet,
http://nssdc.gsfc.nasa.gov/planetary/factsheet/marsfact.html (accessed August 2008)

Figure 1

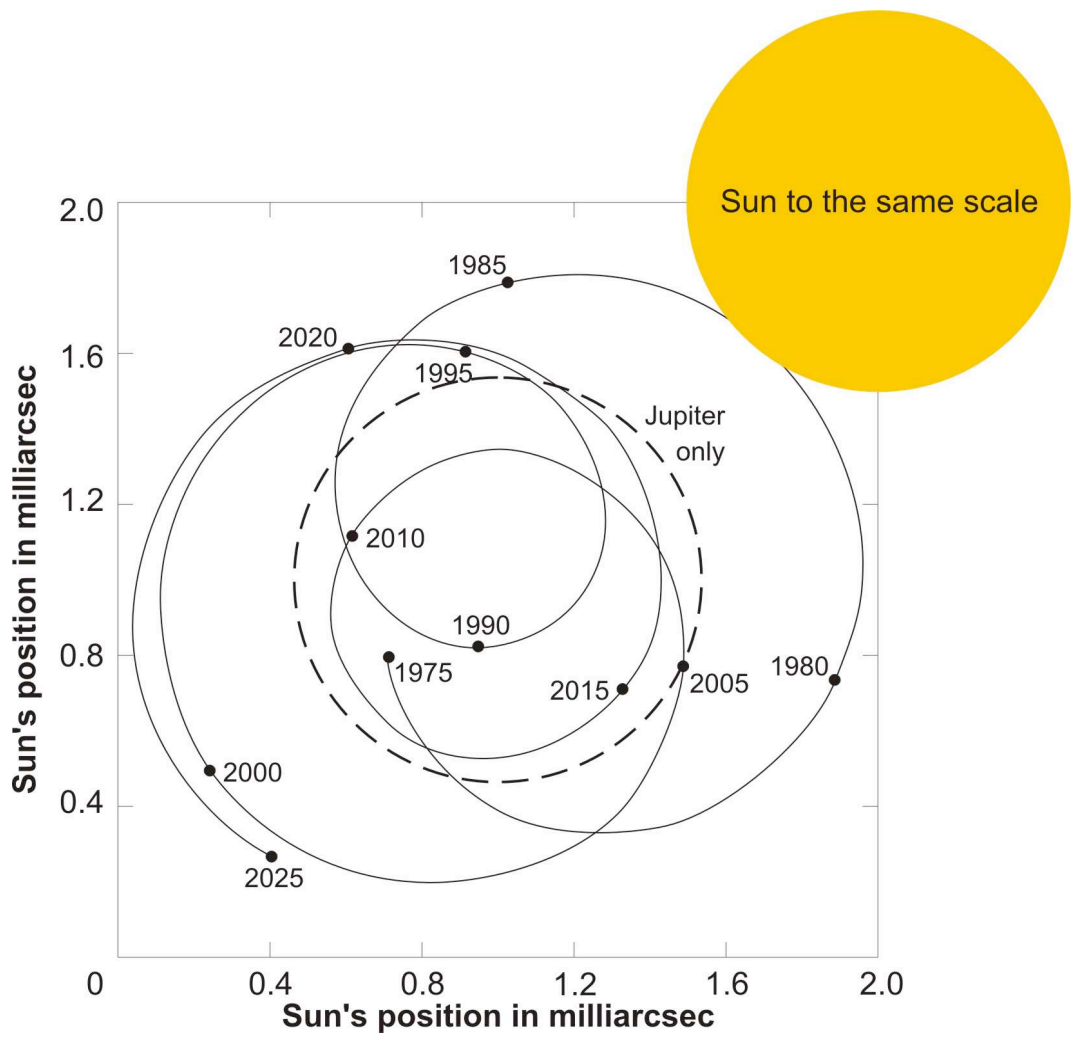

Figure 2

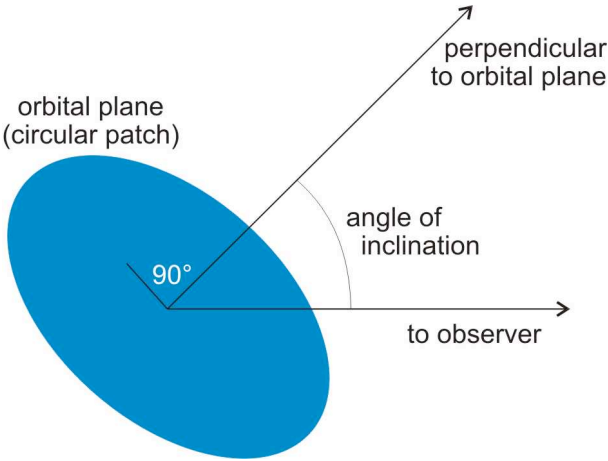

Figure 3

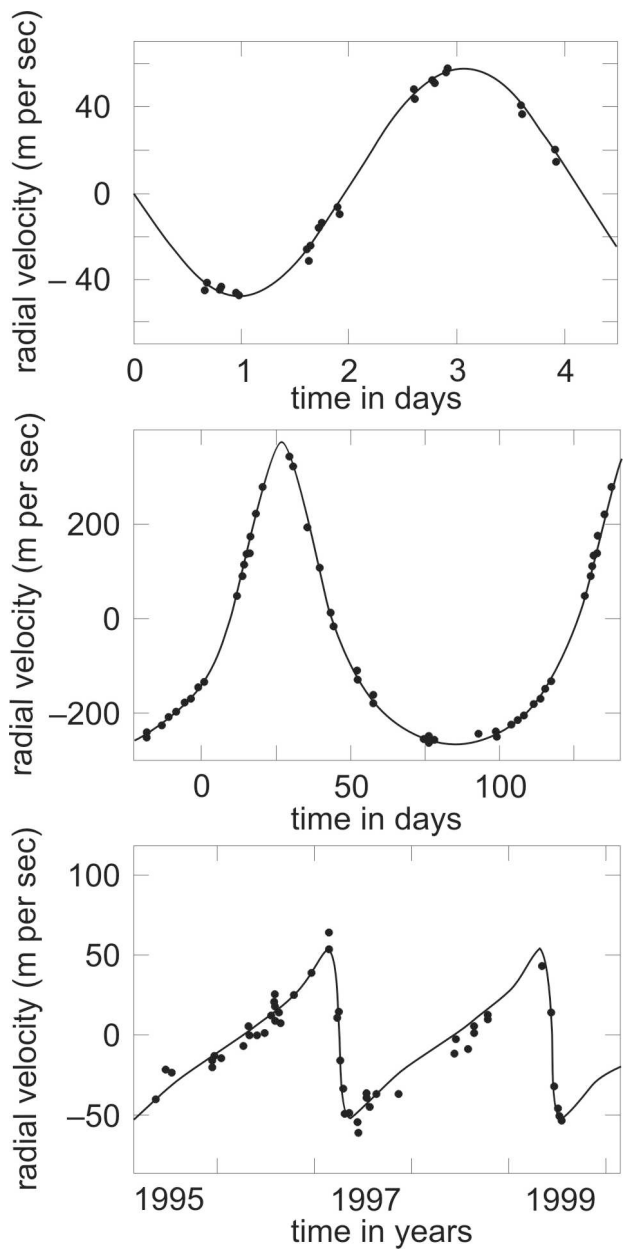

Figure 4

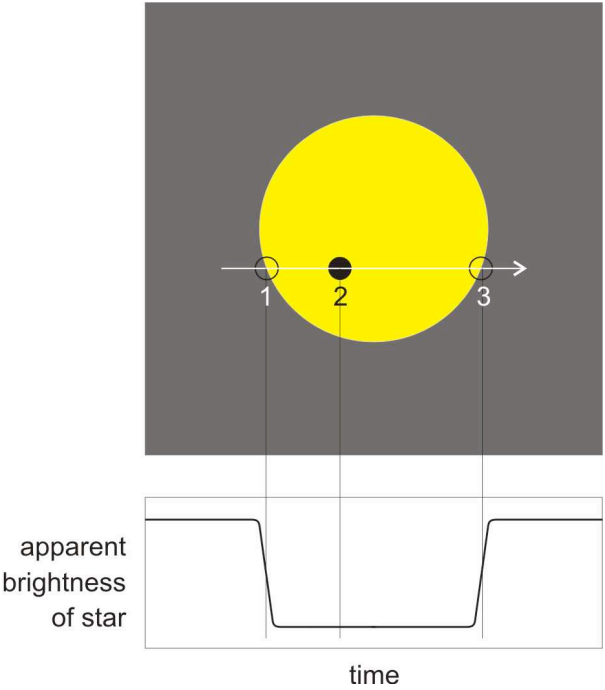

Figure 5

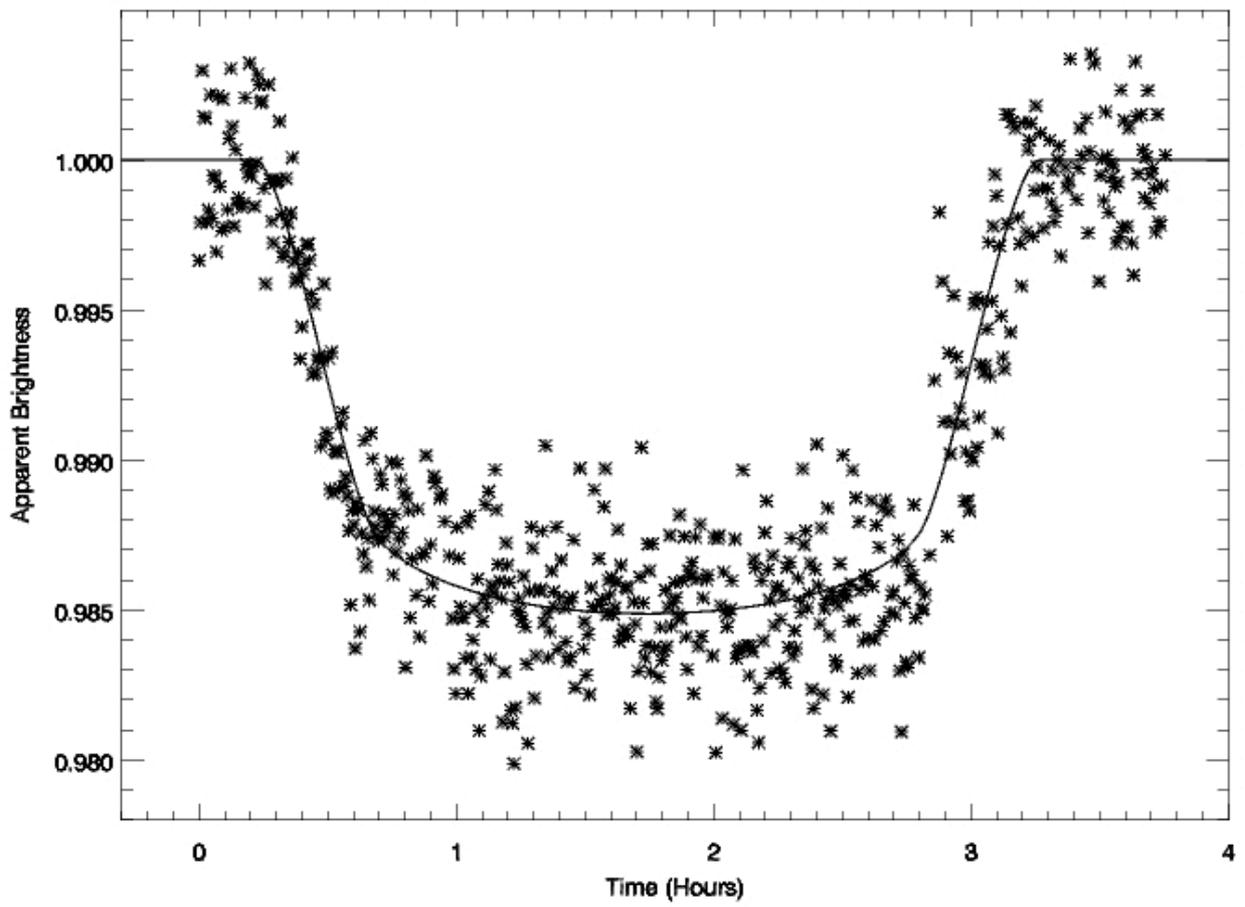

Figure 6

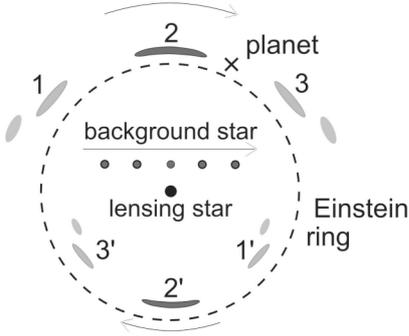 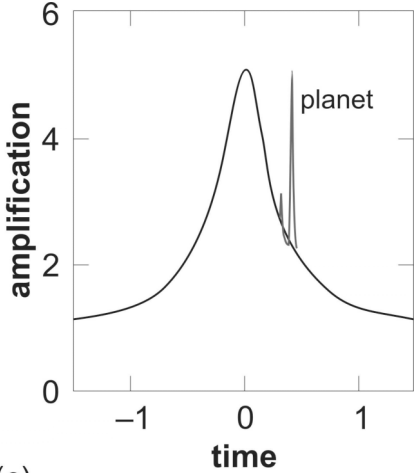

(a)           (c)

Figure 7

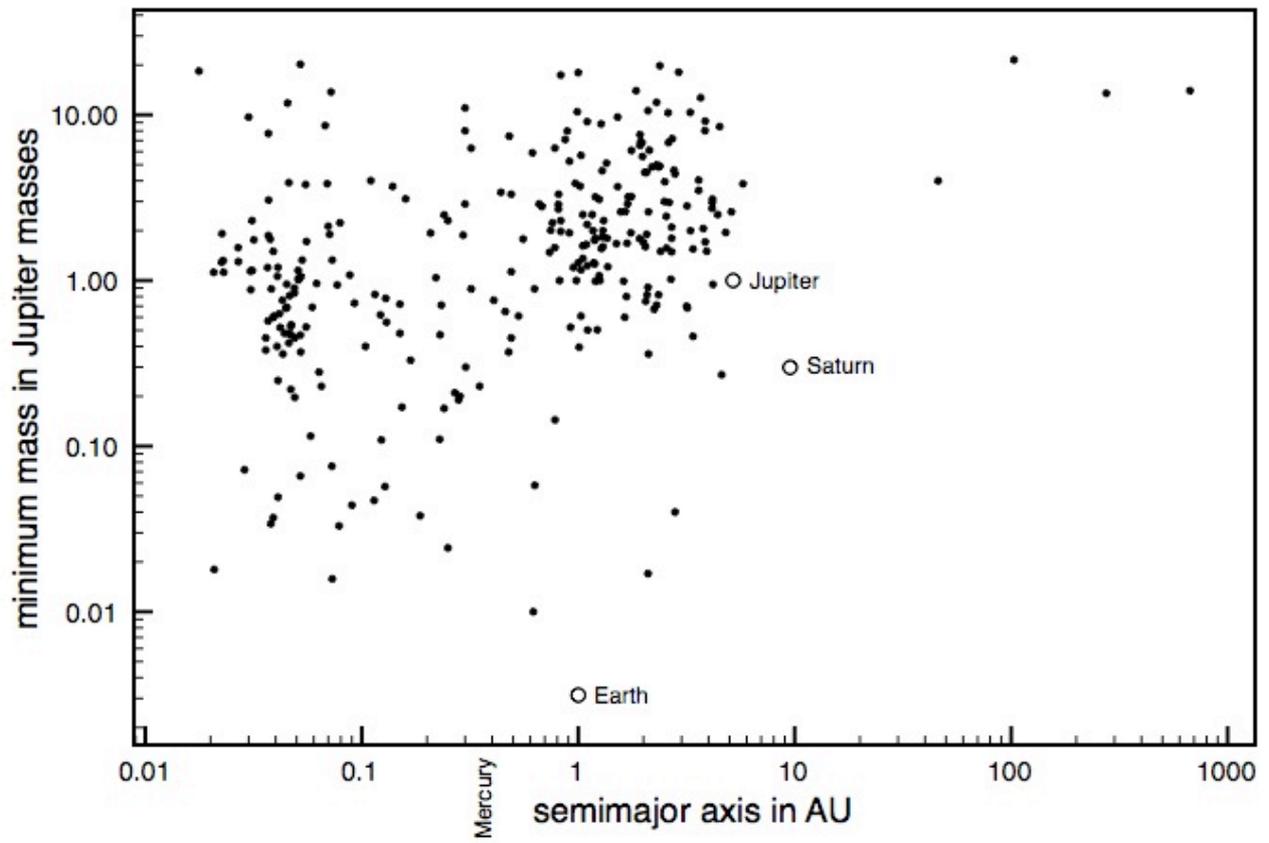

Figure 8

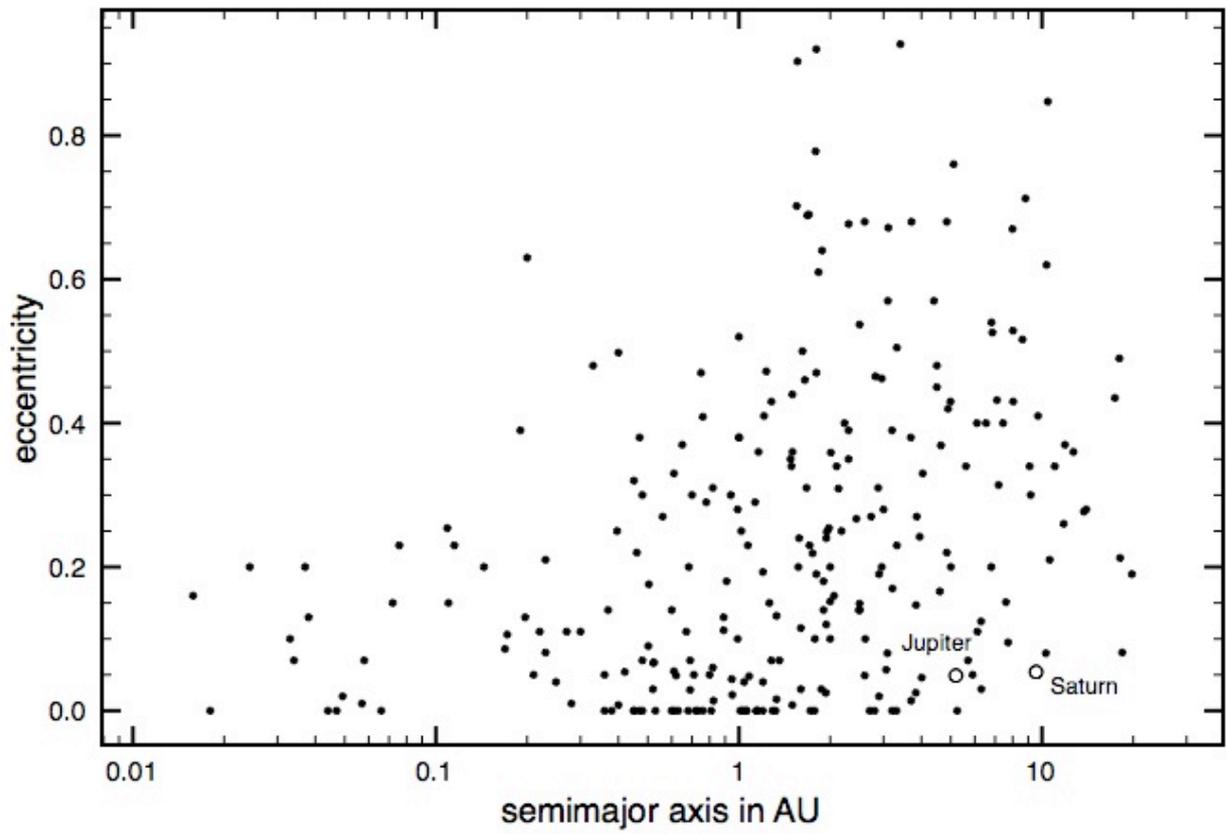

Figure 9

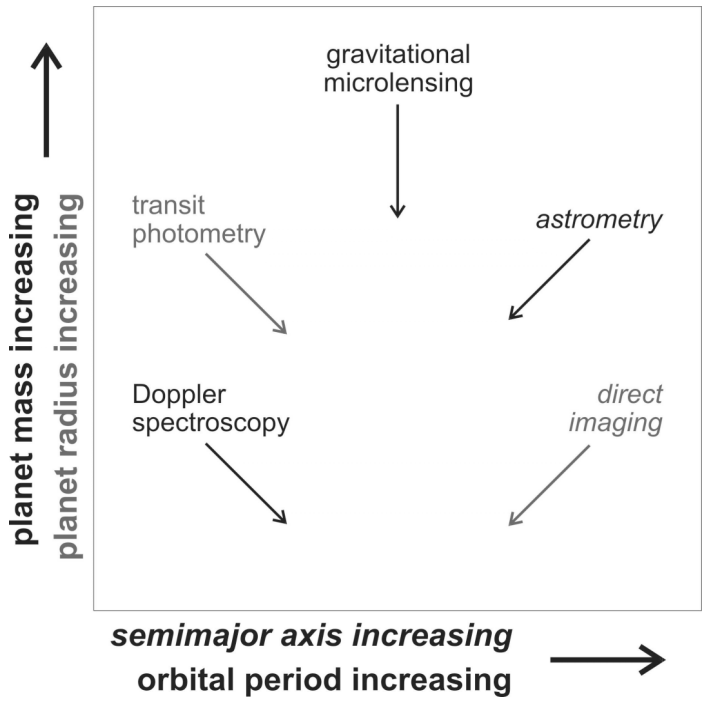

Figure 10

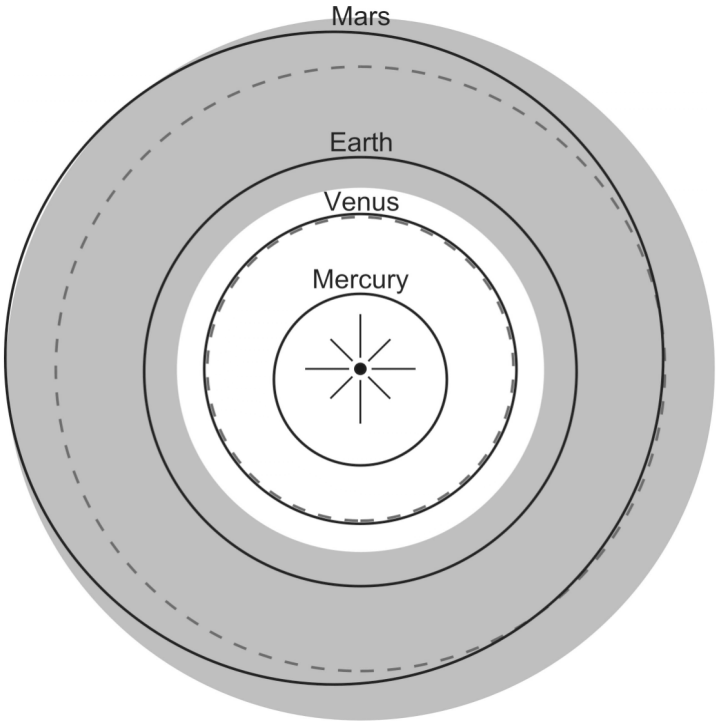

Figure 11

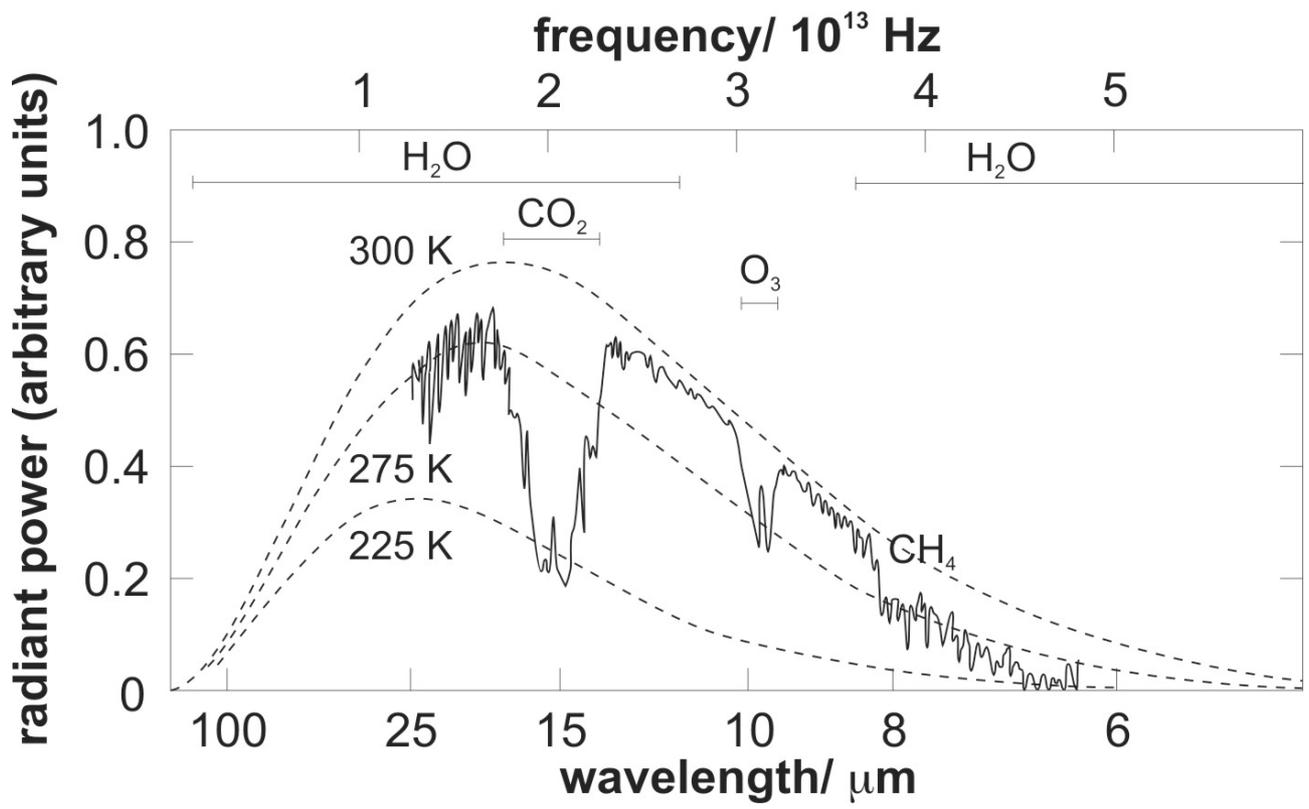

Figure 12

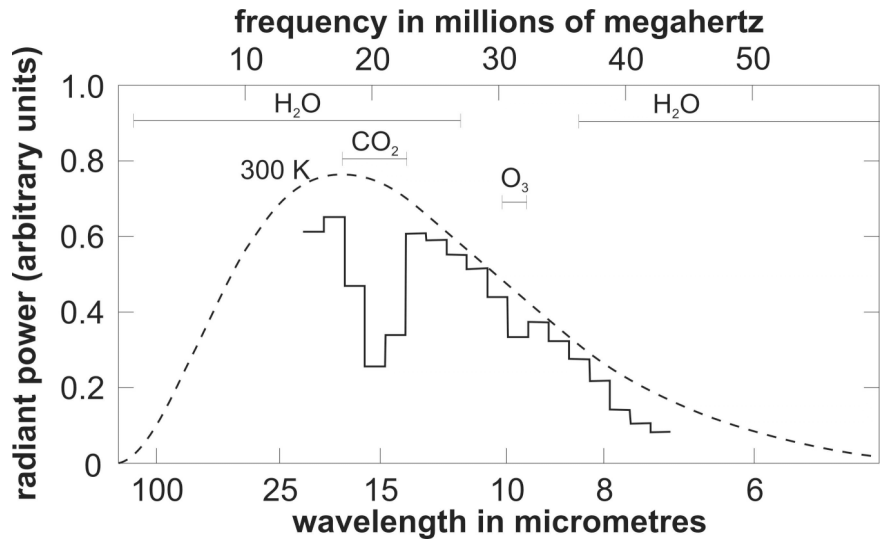